\documentclass[%
 reprint,
superscriptaddress,
 amsmath,amssymb,
 aps,
]{revtex4-2}

\usepackage{graphicx}
\usepackage{dcolumn}
\usepackage{bm}
\usepackage{physics}

\begin{document}

\preprint{APS/123-QED}

\title{Improved constraints on axion inflation with the new code \textmd{\textsc{Axiogen}}: an efficient generator of curvature perturbations}

\author{Matthew Reinhard}
\email{Matt.Reinhard@ufl.edu}
 \affiliation{Department of Physics, University of Florida, Gainesville, FL 32611, USA }
\author{Zachary Slepian}%
 \email{zslepian@ufl.edu}
\affiliation{
 Department of Astronomy, University of Florida, Gainesville, FL 32611, USA
}

\author{Atsushi Taruya}%
 \email{ataruya@yukawa.kyoto-u.ac.jp}
\affiliation{Center for Gravitational Physics and Quantum Information, Yukawa Institute for Theoretical Physics, Kyoto University, Kyoto 606-8502, Japan
}
\affiliation{Korea Institute for Advanced Study, 85 Hoegiro, Dongaemun-gu, Seoul 02455, Republic of Korea}
\affiliation{Kavli Institute for the Physics and Mathematics of the Universe, Todai Institute for Advanced Study, The University of Tokyo (Kavli IPMU, WPI), Kashiwa, Chiba 277-8583, Japan}

\date{\today}

\begin{abstract}
Axion inflation, in which a pseudoscalar inflaton couples to a $U(1)$ gauge field via a Chern-Simons interaction, has garnered significant interest due to its rich phenomenology. The power spectrum and bispectrum for this model are often parametrized by functions known as $f_2$ and $f_3$, respectively. Previous work computed $f_2$ and $f_3$ using an approximation for the gauge field mode functions. While these approximate solutions are fairly accurate ($\mathcal{O}(10\%)$) for large values of the model parameter $\xi$, which controls gauge field particle production, our recent work has shown them to be quite inaccurate for smaller values of $\xi$. We present new fitting functions for $f_2$ and $f_3$ that are significantly more accurate than those previously used. We use these functions to place new constraints on $\xi$ and also set lower bounds for the energy scale of the axion gauge field coupling for $\phi^{1/3}$ and $\phi^{2/3}$ inflationary potentials. Additionally, we present a new public code to efficiently generate realizations of the primordial curvature perturbation for axion inflation in the weak backreaction regime. This code enables numerical computation of higher order correlators and provides a novel path to initial conditions for large-scale structure formation. 
\end{abstract}

\maketitle


\section{Introduction}

The statistics of the cosmic microwave background (CMB) temperature and polarization anisotropies have been successfully reproduced by a fluctuating scalar field in the early universe during a period of rapid expansion known as inflation. This scalar degree of freedom, the inflaton, sources scalar fluctuations in the metric which freeze on superhorizon scales and provide the seeds from which large-scale structure (LSS) formed. 

We focus here on axion inflation, in which a pseudoscalar inflaton field couples to a $U(1)$ gauge field via a Chern-Simons interaction. This model has a rich phenomenology: it can produce large non-Gaussianity \cite{LargeNG, Barnaby_2011, Jamieson_2025}, primordial magnetic fields \cite{Garretson_1992, Anber_2006, Fujita_2015, Adshead_2016}, primordial black holes \cite{Bugaev2013, Linde_2013}, and gravitational waves \cite{Sorbo_2011, Niu_2023, Corba_2026}. 

In this model, gauge field particle production is controlled by the parameter $\xi$, which depends on the motion of the inflaton background and on the axion-gauge field interaction strength. The gauge field is analogous to the photon and can be decomposed into left and right-handed circular polarization states. The axion-gauge field interaction introduces a tachyonic instability, producing copious amounts of gauge quanta with one polarization type while suppressing the other. This process dynamically breaks parity symmetry.

The phenomenological predictions of axion inflation are often difficult to determine analytically. An approximation for the gauge-field mode functions first presented in \cite{Anber_2006} has been widely used to make these calculations tractable \cite{Anber_2010, Barnaby_2011_2, Barnaby_2011,  Sorbo_2011, Barnaby_2012, Cook_2012, Anber_2012, Linde_2013, Meerburg_2013, Ferreira_2014, Garc_a_Bellido_2016, Domcke_2017, _zsoy_2023, alam2024, ferreira2026, Corba_2026}.\\ 
\indent \cite{Barnaby_2011} explicitly demonstrated how to compute the power spectrum and bispectrum for this model, defining the functions $f_2$ and $f_3$ to parameterize these two correlators, respectively. Recently, in \cite{reinhard2026_2}, we presented a new method for computing realizations of the inverse decay curvature perturbation in axion inflation for the case of weak backreaction. In that work, we used these realizations to recompute $f_2$ and $f_3$ for several different values of $\xi$, and found that our results were quite different from \cite{Barnaby_2011}. This was due to \cite{Barnaby_2011}'s using an approximate gauge-field mode function, as the discrepancy disappeared when we redid our analysis using this same approximation. \\
\indent\cite{Durrer_2024} performed a more general computation of the power spectrum and bispectrum, including both gauge-field helicities, metric perturbations, and the real and imaginary parts of the gauge-field mode functions. They compared their results with \cite{Barnaby_2011} for only large values of $\xi$ and found an $O(10\%)$ difference. In \cite{reinhard2026_2}, we found that our results for $f_2$ and $f_3$ are in good agreement with those of \cite{Durrer_2024} for the values of $\xi$ that we tested. \\
\indent In this Letter, we recompute the power spectrum and equilateral bispectrum of the inverse decay curvature perturbations in axion inflation for $1.8 \leq \xi \leq 3$, extending our work \cite{reinhard2026_2}. We then give new fitting functions for $f_2$ and $f_3$. We use \textit{Planck} constraints on the scalar spectral amplitude $A_s$ and the (equilateral-type) nonlinearity parameter $f^{\rm equil}_{NL}$ to place bounds on $\xi$ and the energy scale of the axion-gauge field interaction $\Lambda$. Furthermore, we present a new code, $\textsc{Axiogen}$, for generating primordial curvature perturbations for axion inflation in the weak backreaction regime. $\textsc{Axiogen}$ is publicly available at https://github.com/MattReinhard/Axiogen.

\section{Physical Setup}
We begin by assuming a homogeneous and isotropic Friedmann--Lema\^{i}tre--Robertson–Walker (FLRW) spacetime geometry:
\begin{align}
ds^2 = -dt^2 + a^2(t)d\vb x^2 = a^2(\tau)[- d\tau^2 + d\vb x^2],
\end{align}
where we have set $c \equiv 1$. $a(t)$ is the scale factor and $\tau$ is conformal time, related to the cosmic time $t$ via $dt = a(\tau)d\tau$. We consider axion-$U(1)$ inflation, in which the pseudoscalar inflaton $\phi$ couples to a $U(1)$ gauge field $A^\mu$ via the Chern-Simons interaction with Lagrangian density
\begin{align}
    \mathcal{L}_{\rm int} = -\frac{\sqrt{-g}}{4\Lambda}\,\phi\, F_{\mu \nu} \tilde{F}^{\mu \nu}.
\end{align}
Here $g$ is the metric determinant, $\Lambda^{-1}$ is the axion gauge field coupling, $F_{\mu \nu} \equiv \partial_\mu A_\nu - \partial_\nu A_\mu$ is the field strength tensor, $\tilde F^{\mu \nu} \equiv \epsilon^{\mu \nu \rho \sigma} F_{\rho \sigma}/2\sqrt{-g}$ is its dual, and $\epsilon^{\mu \nu \rho \sigma}$ is the Levi-Civita symbol. 

Here we present only the definitions and equations absolutely necessary for the reader to follow the current work. A more detailed presentation of this model is in \cite{Barnaby_2011}.
We work in Coulomb gauge, setting $A^0 = 0$ and $\div \vb{A} = 0$, and write the gauge field as an inverse Fourier transform (FT):
\begin{align}
    \label{A:ansatz}
    \vb A(\tau, \vb x) = \sum_{\lambda = \pm } \int \frac{\dd[3] \vb k}{(2\pi)^{3}} \bigg [ \bm{\hat{\epsilon}}_\lambda (\vu k) a_\lambda (\vb k) A_\lambda (\tau, k) e^{-i \vb k \vdot \vb x} + \text{h.c.} \bigg ],
\end{align}
where $\pm$ are the two possible polarizations of the gauge field, the circular polarization vectors $\bm{\hat{\epsilon}}_\lambda (\vu k)$ satisfy 
\begin{align}
 \vb k \cdot \bm{\hat{\epsilon}}_{\pm} (\vu k) &= 0, \hspace{2mm} \vb k \times \bm{\hat \epsilon}_{\pm}(\vu k) = \mp ik\bm{\hat \epsilon}_{\pm}(\vu k), \\
& \text{and} \hspace{4mm} \bm{\hat\epsilon}_{\pm}(-\vu k) = \bm{\hat\epsilon}_{\pm}^*(\vu k),
\label{pol_vec}
\end{align}
h.c. means ``Hermitian conjugate", and $*$ means a complex conjugate. $a_\lambda (\vb k)$ and $A_\lambda (\tau, k)$ are the gauge-field annihilation operator and mode function, respectively.

The gauge field experiences a tachyonic instability, resulting in large particle production for the $A_-$ modes \cite{Anber_2010, LargeNG}. Additionally, the real part of the mode function dominates over the imaginary part. We therefore neglect the $A_+$ modes and include only the real part of the gauge-field mode function in what follows.

The mode function is
\begin{align} \label{A_m}
    A_{-}(\tau, k) = \frac{1}{\sqrt{2k}}\, G_0(\xi, -k\tau),
\end{align}
where $G_0(\xi, -k\tau)$ is the Coulomb function of the second kind \footnote{https://mathworld.wolfram.com/CoulombWaveFunction.html}. The parameter $\xi \equiv |\dot{\bar \phi}|/(2\Lambda H)$ controls gauge-field particle production, where $\bar \phi$ is the background value of the inflaton and $H$ is the Hubble parameter. A very commonly-used approximation \cite{Anber_2010, Barnaby_2011_2, Barnaby_2011,  Sorbo_2011, Barnaby_2012, Cook_2012, Anber_2012, Linde_2013, Meerburg_2013, Ferreira_2014, Garc_a_Bellido_2016, Domcke_2017, _zsoy_2023, alam2024, Kume_2025, ferreira2026, Corba_2026} for this mode function is
\begin{align} \label{A_approx}
    A^{\rm approx}_{-}(\tau, k) \approx \frac{1}{\sqrt{2k}}\left(\frac{-k\tau}{2\xi}\right)^{1/4} e^{\pi \xi - 2\sqrt{-2\xi k\tau}}.
\end{align}
In Section \ref{Section3}, we will see that these approximate mode functions lead to significant underestimates of the power spectrum and bispectrum.

\section{Recomputing $f_2$ and $f_3$} \label{Section3}
In axion inflation, the curvature perturbation $\zeta$ receives a contribution from both vacuum fluctuations, and from the inverse decay of gauge quanta into inflaton fluctuations \cite{LargeNG}:
\begin{align}
    \zeta = \zeta^{\rm vac} + \zeta^{\rm inv}.
\end{align}
The inverse decay perturbation $\zeta^{\rm inv}$ can be written as
\begin{align}\label{zeta_inv}
    \zeta^{\rm inv}(\tau, \vb k) &= 8\pi^2\, \xi \, \mathcal{P} \,\bigg(\frac{k}{k_0}\bigg)^{(n_{\rm s} - 1)/2}   \, R(\tau, \vb k),
\end{align}
 where $\mathcal{P}$ is the vacuum power at the pivot scale $k_0$, $n_{\rm s}$ is the scalar spectral index, and $R(\tau,k)$ encodes the response of the inflaton field to the gauge-field source. Our work \cite{reinhard2026_2} presented a method to generate realizations of the response $R$ on a discrete Fourier grid and used these realizations to compute $f_2(\xi)$ for $\xi = 2, 2.5,$ and $3$.

 Generation of the response $R$ involves integrals over conformal time $\tau$ and momentum $\vb q$. In \cite{reinhard2026_2} we determined the proper sampling rate for time integration to ensure numerical convergence to within $1\%$ for $\xi \in \{2,2.5,3\}$. We found that significantly higher sampling was required for convergence as we decreased the value of $\xi$. In Appendix \ref{AppendixA} we demonstrate convergence for $\xi=1.8$.

Here we use our method \cite{reinhard2026_2} to compute $f_2(\xi)$ and $f_3(\xi)$ for values of $\xi$ between $1.8$ and $3$ in increments of $\Delta\xi = 0.05$. We take a brute-force approach by using only the established convergent sampling rates for $\xi \in \{ 1.8,2,2.5,3\}$. As convergence requires finer time integral sampling for smaller values of $\xi$, we use the established convergent settings for the bottom of each range. For example, we use the convergent sampling rate for $\xi=1.8$ when generating realizations with $\xi \in \{1.8,1.85, 1.9, 1.95\}$, and use the convergent sampling rate for $\xi=2$ for realizations with $\xi$ between $2$ and $2.45$. This approach uses finer sampling than is required for $\mathcal{O}(1\%)$ convergence, increasing the time required to generate realizations, but avoids the need to determine $\mathcal{O}(1\%)$ convergence settings for each value of $\xi$. 
 
\subsection{The Power Spectrum} \label{Sec_PS}

The inverse-decay dimensionless power spectrum is 
\begin{align}
    \Delta_{\zeta}^{2 \; \rm inv}(k) = \bigg(\frac{k}{k_0}\bigg)^{n_{\rm s} - 1} \mathcal{P}^2 f_2(\xi) \, e^{4\pi\xi},
\end{align}
where $f_2(\xi)$ was first defined in \cite{Barnaby_2011} and involves an integral over gauge-field mode functions.   
 
\begin{figure*}[t]
    \centering
    \includegraphics[width=1\textwidth]{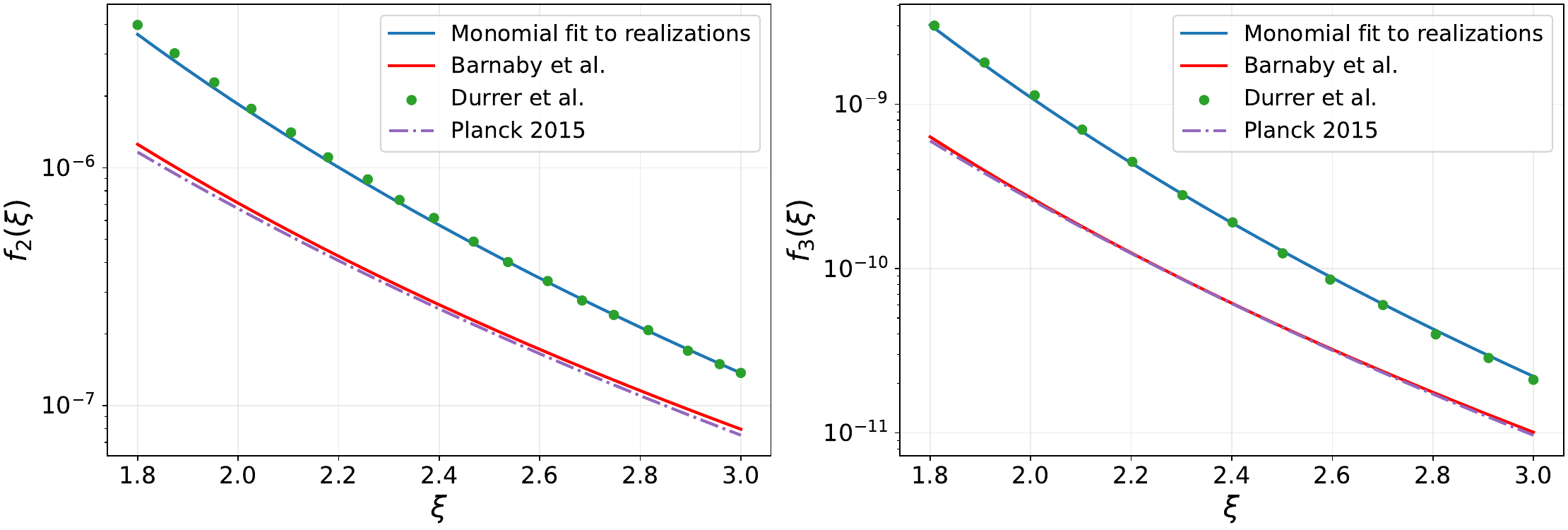}
    \caption{The fits for functions $f_2(\xi)$ (left) and $f_{3}(\xi)$ (right) which were computed from realizations of the stochastic response $R$ (as described in Section \ref{Section3}) are shown. Also shown are results reproduced from Durrer et al. \cite{Durrer_2024}, Barnaby et al. \cite{Barnaby_2011}, and the \textit{Planck} collaboration \cite{Planck_2016}. We find that our results are in good agreement with Durrer et al., but differ significantly from Barnaby et al. and \cite{Planck_2016}. This difference is due to the fact that we use the exact mode functions (\ref{A_m}), whereas Barnaby et al. and (likely) \cite{Planck_2016} use the approximate mode functions (\ref{A_approx}), leading to a significant underestimate of the power spectrum and bispectrum.}
    \label{fig:f2_andf_3}
\end{figure*}

 For each value of $\xi$ we generate 20 realizations of the response $R$, compute the dimensionless power spectrum
\begin{align}\label{PS_def}
  \Delta^2_{R}(\kappa) \equiv \frac{\kappa^3}{2\pi^2}\, P_{R}(\kappa)    
\end{align}
  for each, and average them. Here $\kappa$ is a dimensionless wavenumber and $P_{R}(\kappa)$ is the power spectrum of $R$. We then average this result for $\Delta^2_{R}(\kappa)$ over the convergent range $10 \leq \kappa \leq 30$ (\cite{reinhard2026_2} gives further details). Denoting the result of this average $\bar{\Delta}^2_{R}$, we compute $f_2(\xi)$ via
\begin{align}\label{f2}
    f_{2}(\xi) = (8\pi^2\xi)^2 \,e^{-4\pi\xi}\, \bar{\Delta}_{R}^{2}. 
\end{align}
We then fit a monomial to our measured values of $f_{2}(\xi)$, obtaining
\begin{align} \label{f2_fit}
    f_2(\xi) = \frac{1.59 \times10^{-4}}{\xi^{6.42}}.
\end{align}
\cite{Durrer_2024} calculated $f_2(\xi)$ taking into account both gauge-field polarizations, both real and imaginary parts of the mode functions, and metric perturbations. We used the \textsc{PlotDigitizer} \footnote{https://plotdigitizer.com/} tool to extract these points (\cite{Durrer_2024} Figure 1a). These points and the fit to our own realizations, Eq. (\ref{f2_fit}), are shown in Figure \ref{fig:f2_andf_3}. We find that our result for $f_2(\xi)$ closely matches those of \cite{Durrer_2024}.

\cite{Barnaby_2011} neglects the suppressed gauge-field modes and includes only the real part of the mode functions (as we do in this work); however, they use the approximate mode function (\ref{A_approx}) for their computation of $f_2(\xi)$. They give the fit
\begin{align}
    f^{\rm approx}_2(\xi) = \frac{3 \times10^{-5}}{\xi^{5.4}}.
\end{align}
We include this, and also the fit for $f_2(\xi)$ used by the \textit{Planck} collaboration \cite{Planck_2016} for their constraints on this inflationary model, in Figure \ref{fig:f2_andf_3}. \cite{Planck_2016} does not explicitly state whether they used the exact (Eq. \ref{A_m}) or approximate (Eq. \ref{A_approx}) mode functions to compute $f_2$, but their fit closely matches that of \cite{Barnaby_2011}, suggesting use of the approximate mode functions. 

\subsection{The Bispectrum}
Following \cite{Barnaby_2011, reinhard2026_2}, we can write the inverse decay bispectrum as
\begin{align} \label{Bispectrum_eqn}
B_{\zeta}^{\rm inv}(k_1, k_2, k_3) = 
\frac{3}{10}(2\pi)^{4} \, \mathcal{P}^3\, e^{6\pi\xi}\,
\frac{1+x_2^3+x_3^3}{k^6 x_2^3 x_3^3}\,
f_3(\xi,x_2,x_3), 
\end{align}
where $|\vb k_1| \equiv k$, $|\vb k_2| \equiv x_2 \, k$, and $|\vb k_3| \equiv x_3 \, k$. We compute $f_3(\xi) \equiv f_3(\xi,1,1)$ for the equilateral case via 
\begin{align}\label{f3_BR}
    f_3(\xi, 1, 1) = \frac{1280}{9}\, \pi^6\, \xi^3 \,e^{-6\pi\xi} \,\tilde{B}_R(\kappa ,\kappa ,\kappa),
\end{align}
where $\tilde{B}_R(\kappa ,\kappa ,\kappa)$ is the dimensionless equilateral bispectrum of the response $R$. At each value of $\xi$, $\tilde{B}_R(\kappa ,\kappa ,\kappa)$ was measured at $\kappa=20$ from $100$ realizations of $R$ and then averaged. We chose this value of $\kappa$ as it sits safely in the converged region of the power spectrum (see \cite{reinhard2026_2} and Appendix \ref{AppendixA} for more details). We then fit a monomial to our measured values of $f_3(\xi)$, finding
\begin{align}\label{f3_fit}
    f_3(\xi) = \frac{8.78\times10^{-7}}{\xi^{9.64}}.
\end{align}
\cite{Barnaby_2011} found a fit to $f_3(\xi)$ using the approximate mode function: 
\begin{align}
    f_3^{\rm approx}(\xi) = \frac{7.4\times10^{-8}}{\xi^{8.1}}.
\end{align}
In the right panel of Figure \ref{fig:f2_andf_3} we plot these functions. We also include points extracted from the \cite{Durrer_2024} Figure 2a, as well as the fit used in the \textit{Planck} analysis \cite{Planck_2016}. We again find that our fit $f_3(\xi)$ is in very good agreement with \cite{Durrer_2024} and $f_3^{\rm approx}(\xi)$ closely matches \cite{Planck_2016}, but there is a large discrepancy between $f_3(\xi)$ and $f_3^{\rm approx}(\xi)$.

\section{Axiogen}
\textsc{Axiogen} is our new code to generate primordial curvature perturbations for axion inflation in the weak backreaction regime. Our code generates realizations of both the vacuum and inverse-decay curvature perturbations on a discrete Fourier grid of side length $L$ and grid size $N$. 

Many of the calculations are done in dimensionless coordinates:
\begin{align}
    \boldsymbol{\kappa} = \frac{\vb k}{k_*},  \qquad  \tilde L = k_* L, \qquad \tilde{\tau} = k_* \tau
\end{align}
where $\vb{k}$ is a wave-vector, $\boldsymbol {\kappa}$ is a dimensionless wave-vector, $\tilde{L}$ is the dimensionless box length, and $k_*$ is a quantity with units $h \, \rm{Mpc}^{-1}$ that sets the conversion between dimensionless and dimensionful coordinates. The default settings for the dimensionless box geometry are $\tilde{L} = 2\pi$ and $N = 256$. These settings were demonstrated to have a convergent inverse decay power spectrum for $10 \lesssim \kappa \lesssim 40$ with $1.8 \leq \xi \leq 3$ in \cite{reinhard2026_2} and Appendix \ref{AppendixA} (for the case $\xi = 1.8$).

The code uses a default conversion factor of $k_* = 2\pi/2000 \, h \, \rm{Mpc}^{-1}$ which yields a box length of $L = 2000 \, h^{-1} \, \rm{Mpc}$ matching the size of the \texttt{Abacus} Dark Energy Spectroscopic Instrument (DESI) Year 1 (Y1) Luminous Red Galaxy (LRG) mock catalogs. This choice of $k_*$ corresponds to a physically converged scale of $0.0314 \, h \,{\rm Mpc}^{-1} \lesssim k \lesssim 0.126 \, h \, {\rm Mpc}^{-1}$, which lies within the range explored in the DESI Y1 LRG full-shape analysis \cite{Adame_2025}. 

 In \cite{reinhard2026_2}, we presented an algorithm to generate realizations of the inverse decay curvature perturbation. In Section \ref{Section3} we used this algorithm to determine $f_2(\xi)$, which parametrizes the power of these inverse decay fluctuations for $1.8 \leq \xi \leq 3$. 
 
 The dimensionless power spectrum, including vacuum and inverse decay effects, is 
\begin{align}
\label{PS}
    \Delta_{\zeta}^{2}(k) = \mathcal{P}\bigg(\frac{k}{k_0}\bigg)^{n_{\rm s}-1}\Bigg[1 + \mathcal{P} \,f_{2}(\xi)\,e^{4\pi \xi}\Bigg].
\end{align}
Solving for $\mathcal{P}$ the quadratic equation that stems from expanding out Eq. (\ref{PS}), we find 
\begin{align}\label{vacuum_power}
    \mathcal{P}(\xi) = \frac{e^{-4\pi \xi}}{2f_2(\xi)}\Bigg[-1 + \sqrt{1 + 4f_2(\xi)\,e^{4\pi \xi}\,\Delta_{\zeta}^{2}(k_0)}\Bigg].
\end{align}

The \textit{Planck} collaboration reports $10^9A_s = 2.100 \pm 0.030$ at the pivot scale $k_0 = 0.05 \; \rm{Mpc}^{-1}$ \cite{Planck_2020}. \textsc{Axiogen} uses Eq. (\ref{vacuum_power}) along with the fit for $f_2(\xi)$ given in Eq. (\ref{f2_fit}) to determine the spectral amplitude at the pivot scale for the input value of $\xi$.

The vacuum curvature perturbations are generated by first initializing white noise at each point of a dimensionless coordinate grid with volume and grid spacing
\begin{align}
    \tilde{V} = \tilde{L}^3 \qquad \text{and} \qquad \Delta \tilde{x} = \frac{\tilde{L}}{N}, 
\end{align}
respectively. The values of this white noise field $X(\tilde{\vb x})$ are drawn at each point from a Gaussian distribution with zero mean and a variance of $\sigma^2 = (\Delta \tilde{x})^{-3}$ \footnote{this corresponds to a delta function normalization in the continuum}. \textsc{Axiogen} then takes the FT of $X(\tilde{\vb x})$ and rescales the result to obtain the dimensionless vacuum curvature perturbation
\begin{align}
\tilde{ \zeta}^{\rm vac}(\boldsymbol \kappa) = \sqrt{\frac{2\pi^2}{\kappa^3}\mathcal{P}(\xi)}\, \bigg (\frac{\kappa k_*}{k_0}\bigg)^{(n_{\rm s}-1)/2}\, Y(\boldsymbol \kappa),
\end{align}
where $Y(\boldsymbol \kappa)$ is the FT of $X(\tilde{\vb x})$. This rescaling ensures that these perturbations produce the desired vacuum power spectrum in Eq. (\ref{PS}). We also set the Fourier coefficient of the zero mode to zero; this enforces that the average of our perturbations vanishes.

\textsc{Axiogen} separately produces the dimensionless inverse decay perturbation
\begin{align}
    \tilde{\zeta}^{\rm inv}(\tau, \boldsymbol \kappa) &= 8\pi^2\, \xi \, \mathcal{P}(\xi) \,\bigg(\frac{\kappa k_*}{k_0}\bigg)^{(n_{\rm s} - 1)/2}   \, \tilde{R}(\tau, \boldsymbol \kappa),
\end{align}
which is just the dimensionless form of Eq. (\ref{zeta_inv}). \textsc{Axiogen} uses our algorithm \cite{reinhard2026_2} to compute the dimensionless response $\tilde{R}(\tau, \boldsymbol \kappa)$. It then outputs the physical perturbations by multiplying the dimensionless curvature perturbations by $k_*^{-3}$.

\section{Constraints on Non-Gaussianity}
The bispectrum was shown to be nearly equilateral in \cite{Barnaby_2011, Durrer_2024}. We therefore compute the equilateral nonlinearity parameter \cite{Barnaby_2011}
\begin{align}\label{f_NL}
    f_{\rm NL}^{\rm equil} \equiv \frac{f_3(\xi)\, \mathcal{P}^3(\xi)\,e^{6\pi \xi}}{[\Delta_{\zeta}^{2}(k_0)]^2}.
\end{align}
 We use Eqs. (\ref{vacuum_power}) and (\ref{f3_fit}) to express Eq. (\ref{f_NL}) as a function of only $\xi$, and then obtain its root using \textsc{Mathematica}. An analysis of \textit{Planck} data found $f_{\rm NL}^{\rm equil} = -26 \pm 47$ (68\% CL) \cite{planck2018}. Since $f_{\rm NL}^{\rm equil} > 0$ for this model, we impose the upper bound of the central $2\sigma$ interval, \textit{i.e.} we take $f_{\rm NL}^{\rm equil} \leq 68$. {\bf From this we find the upper bound that $\xi \leq 2.5$.}

We now consider the monomial family of potentials
\begin{align}
    &V(\phi) \propto \phi^p.
\end{align}

The slow-roll approximation yields \cite{Barnaby_2011}
\begin{align}\label{xi_sr}
   \xi = \frac{M_{\rm p}}{2\Lambda}\sqrt{\frac{p}{2N}}, 
\end{align}
where $M_{\rm p}$ is the reduced Planck mass and $N$ is the number of e-folds between the moment the CMB scales exited the horizon and the end of inflation. We consider the cases of $p=2/3$ with $N=60$ and $p=1/3$ with $N=50$, as these potentials are still viable according to recent CMB measurements \cite{calabrese2025}. Eq. (\ref{xi_sr}) allows us to directly translate our constraint on $\xi$ to a constraint on the energy scale of the axion-gauge field interaction $\Lambda$ for these two cases.  We find $\Lambda \geq 3.6 \times10^{16}$ GeV for $p=2/3$ and $\Lambda \geq 2.8 \times10^{16}$ GeV for $p=1/3$.

\section{Conclusion}
The seminal work of \cite{Barnaby_2011} demonstrated how to compute the power spectrum and bispectrum for axion inflation using the approximate mode functions of \cite{Anber_2006}. Their provided expressions for the functions $f_2(\xi)$ and $f_3(\xi)$ have been used in multiple \textit{Planck} analyses \cite{Planck_2016, Planck_2016_NG, planck2018}. \cite{Durrer_2024} gave a more general framework for this computation, including both the real and imaginary parts of the gauge-field mode functions, metric perturbations, and both polarizations. They compared their results to those of \cite{Barnaby_2011} only at large $\xi$, finding an $\mathcal{O}(10\%)$ difference. In \cite{reinhard2026_2}, we presented a new algorithm for generating realizations of the inverse-decay curvature perturbation and used them to compute $f_2(\xi)$ and $f_3(\xi)$ for several values of $\xi$. Our results differed significantly from those of \cite{Barnaby_2011}, especially for smaller values of $\xi$. In the current work, we have used the algorithm of \cite{reinhard2026_2} to obtain new fitting functions for $f_2(\xi)$ and $f_3(\xi)$ in the range $1.8 \leq \xi \leq 3$. These functions can be used in future CMB analyses to provide more accurate constraints on axion inflation. Furthermore, we used \textit{Planck} constraints on $A_s$ and $f^{\rm equil}_{NL}$ to determine an upper bound of $\xi \leq 2.5$ and a lower bound on the energy scale of the axion-gauge field interaction for both the $\phi^{1/3}$ and $\phi^{2/3}$ potentials. 

In the current work, we have also presented a new code, \textsc{Axiogen}, that generates realizations of both vacuum and inverse-decay curvature perturbations in the weak backreaction regime, with a combined power consistent with \textit{Planck} measurements. These realizations offer a novel path to obtaining initial conditions for LSS formation, and enable numerical computation of higher-order correlators. We plan to use \textsc{Axiogen} to further constrain axion inflation with current and upcoming LSS data, such as DESI, Euclid, Roman, and SPHEREx. We are particularly interested in searching for the signal of this model in the parity-odd galaxy 4-Point Correlation Function.



\appendix 

\section{Convergence Tests for $\xi=1.8$} \label{AppendixA}
Here we closely follow \cite{reinhard2026_2} to find settings such that the integrals required to generate realizations of the response $R(\tau, \vb k)$ are convergent for the case of $\xi = 1.8$. 

\begin{align}\label{R_full}
   R(\tau, \boldsymbol k) =  \int_{-\infty}^0 d\tau' \, \mathcal{K}(\tau, \tau', k) \, S(\tau', \vb k),
\end{align}
where $S(\tau', \vb k)$ is a convolution involving gauge field mode functions. We find it easier to work with the dimensionless response $\tilde{R}(\tilde{\tau}, \boldsymbol \kappa) = k_*^{3} R(\tau, \boldsymbol k)$.

We first test that our integral converges as the late-time bound approaches zero. We compare $\tilde {R}(\tilde \tau, \boldsymbol \kappa)$ against a late-time reference at different values of $\tilde{\tau}$ via the relative $\ell^2$-norm:
\begin{align}
    \text{error} = \frac{\norm{R(\vb k) - R_{\rm ref}(\vb k)}_2}{\norm{R_{\rm ref}(\vb k)}_2}.
\end{align}
For this test, we set our grid size to $N_g = 256^3$, $\tilde{L} = 2\pi$, and set the number of sampling points for time integration to $N_t = 500$. 
We find that the same late-time bound of $|\tilde \tau| = 0.09$ used in \cite{reinhard2026_2} has a relative error of $0.46\%$ when compared against a $|\tilde{\tau}| = 0.0075$ reference. 

We next determine the number of sampling points $N_t$ required for time integration. We again set $N_g = 256^3$, $\tilde{L} = 2\pi$, and fix a late-time integration bound of $|\tilde \tau| = 0.09$. We compared the response generated with different number of sample points for the time integration in (\ref{R_full}) against an $N_t = 500$ reference using the relative $\ell^2$-norm. The results are shown in Table \ref{time-convergence}. 

Informed by these two convergence tests, we choose a late-time bound of $|\tilde \tau| = 0.09$ and a time integration sampling of $N_t = 150$ when producing realizations of  $\tilde{R}(\tilde{\tau}, \boldsymbol \kappa)$ to measure the power spectrum and bispectrum in Section \ref{Section3}. We also use these settings for the remaining convergence tests in this appendix.  

\begin{table}
    \centering
    {\large
    \scalebox{0.92}{%
    \begin{tabular}{|c|c|}
        \multicolumn{2}{c}{$\xi = 1.8$} \\[3pt]
        \hline
        $N_t$ & Percent error \\
        \hline
        120 & 1.126\% \\
        \textbf{150} & \textbf{0.919\%} \\
        180 & 0.793\% \\
        300 & 0.500\% \\
        500 & Reference \\
        \hline
    \end{tabular}%
    }\par
    }
    \caption{Displayed is the dimensionless response $\tilde{R}(\tilde{\tau}, \boldsymbol \kappa)$ computed with different numbers of sampling points $N_t$ for the time integration in Eq. (\ref{R_full}). The errors are calculated by comparing each result against an $N_t = 500$ reference via the relative $\ell^2$-norm. The production setting used for our computation of $f_2$ and $f_3$ in Section \ref{Section3} is shown in bold.}
    \label{time-convergence}
\end{table}

\begin{figure*}[t]
    \centering
    \includegraphics[width=0.6\textwidth]{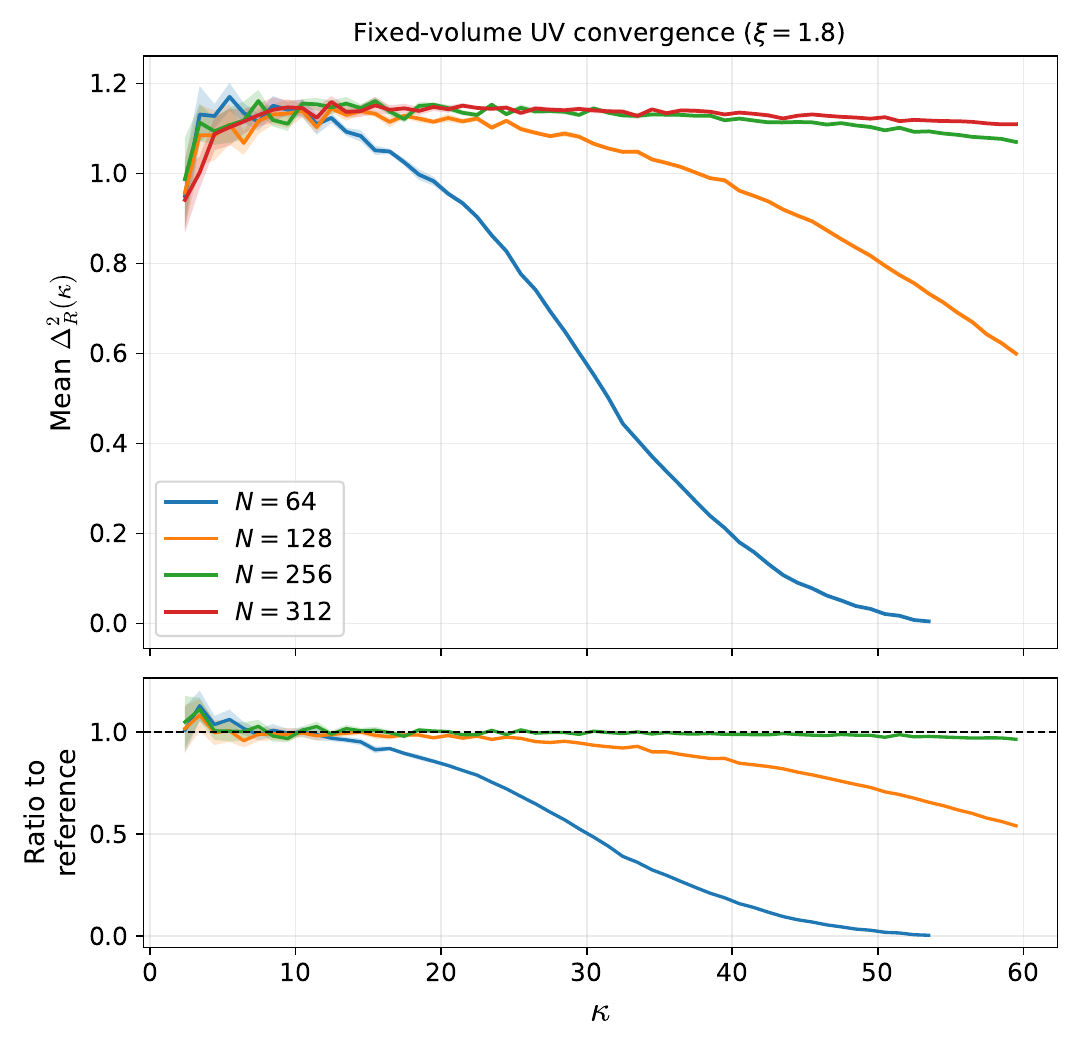}
    \caption{In the top panel, the dimensionless power spectrum of the inverse decay response $\Delta^2_{R}(\kappa)$ is shown for different grid sizes N $\in\{64, 128, 256, 312\}$ for the case of $\xi = 1.8$. For each grid size, 10 realizations were generated and their power spectra were computed using the \texttt{pylians3} library; we then averaged the power spectra at each $\kappa$. The shaded region around each curve gives the $1 \sigma$ uncertainty of the mean. In the bottom panel, the ratio of the power spectra to the N = 312 reference is shown. As the spectrum should be approximately flat, we note that for $N=256$ there is very good convergence in the region $10 \leq \kappa \leq 40$.}
    \label{fig:UV}
\end{figure*}

\begin{figure*}[t]
    \centering
    \includegraphics[width=0.6\textwidth]{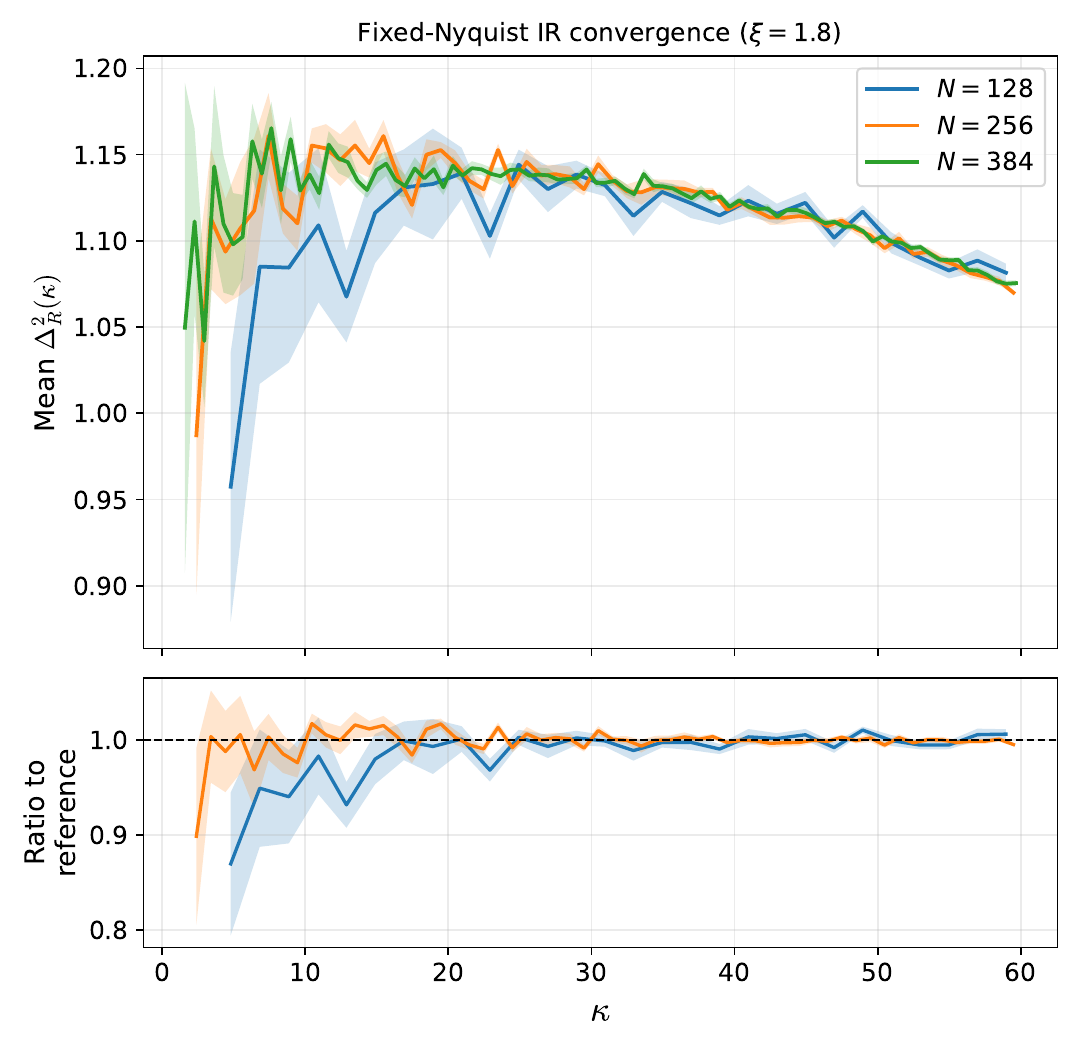}
    \caption{In the top panel, the dimensionless response power spectrum $\Delta^2_{R}(\kappa)$ with $\xi=1.8$ is shown for the different combinations $(\rm{N}, \tilde{L}) \in \{(128,\pi), (256, 2\pi), (384, 3\pi)\}$, which correspond to the grid resolutions $\Delta \kappa \in \{2, 1, \frac{2}{3}\}$, respectively. In the bottom panel, the ratio of each of the two smaller grid power spectra and the $N = 384$ reference are shown. The $N = 256$ and $N = 384$ curves show good agreement for the entire $\kappa$ range, indicating that $S(\tau', \vb k)$ has already converged in the IR for $\kappa_{Ny} = 1$. }
    \label{fig:IR}
\end{figure*}

We next turn to UV (large $\kappa$) and IR (small $\kappa$) convergence. The UV and IR bounds 
\begin{align}
    &\kappa_{Ny} = \frac{N\pi}{\tilde{L}} \hspace{10mm} \text{and}  \hspace{10mm} \kappa_{\rm IR} = \frac{2\pi}{\tilde{L}},
\end{align}
are set by the Nyquist wavenumber and the fundamental mode of our dimensionless grid, respectively. We first fix $\tilde L = 2\pi$ and vary the number of grid points. This fixes the grid resolution $\kappa_{IR}$ while increasing the UV bound for the convolution integral $S(\tau', \vb k)$. We use the dimensionless power spectrum from Eq. (\ref{PS_def}) as a diagnostic for convergence as a converged spectrum should be approximately flat. All spectra were computed using the \texttt{pylians3} library. We find that a grid size of $N = 256$ is sufficiently converged for the region $10 \leq \kappa \leq 40$. The results are shown in Figure \ref{fig:UV}.
For the IR convergence test we consider three different combinations of dimensionless box length $\tilde{L}$ and grid size $N$: $(N, \tilde{L}) \in \{(128,\pi), (256, 2\pi), (384, 3\pi)\}$. We note that these combinations keep the ratio $N/L$ fixed which ensures that the UV bound $\kappa_{Ny}$ is unchanged. The results can be seen in Figure \ref{fig:IR}. We find that the power spectrum is sufficiently converged for the case of $N=256$ and $\tilde{L} = 2\pi$, where the IR bound is $\kappa_{IR} = 1$.   

\begin{acknowledgements}
This work was supported in part by JSPS KAKENHI Grant Numbers 23K25868 and 26H02044 (AT). AT also acknowledges support from FY2025 MUSUBIME of Kyoto University. We acknowledge the use of ChatGPT for code development, but did not use it at all for the writing of this manuscript. ZS acknowledges support from NASA grant number 80NSSC24M0021. We thank the Slepian research group at the University of Florida for useful conversations.
\end{acknowledgements}

\clearpage
\bibliography{bibliography}

\end{document}